\documentclass [10pt] {article}
\usepackage [utf8] {inputenc}
\usepackage[english]{babel}

\usepackage {graphicx}
\usepackage {amsfonts}
\usepackage {amsthm}
\usepackage {amsmath}
\usepackage {natbib}
\usepackage[in]{fullpage} 

\graphicspath{{Figures/}} 

\usepackage{placeins}

\date{ }
\usepackage{color}

\newcommand{\blind}{1}
\newcommand{\papertitle}{
 A simulation study to compare $^{210}$Pb dating data analyses 
}

\begin{document}
	\def\spacingset#1{\renewcommand{\baselinestretch}%
		{#1}\small\normalsize} \spacingset{1}
	\if1\blind
	{
		\title{\textbf{\papertitle}}

		\author{Marco A Aquino-L\'opez\thanks{
				Centro de Investigaci\'on en Matem\'aticas (CIMAT),
				Jalisco s/n, Valenciana, 36023 Guanajuato, Gto, Mexico.
				email: \texttt{aquino@cimat.mx} } \thanks{Corresponding author.}
					\and
			Nicole K. Sanderson\thanks{
				GEOTOP Research Centre, Université du Québec à Montréal, 
				Montréal, Québec, H2X 3Y7, Canada. 
				email: \texttt{sanderson.nicole@uqam.ca}}
					\and
			Maarten Blaauw\thanks{School of Natural and Built Environment,
				Queen's University Belfast,
				Belfast, BT7-1NN, UK.
				email:\texttt{maarten.blaauw@qub.ac.uk}  }
					\and
			Joan-Albert Sanchez-Cabeza\thanks{
				Unidad Acad\'emica de Mazatl\'an, 
				Instituto de Ciencias del Mar y Limnolog\'ia, 
				Universidad Nacional Aut\'onoma de Mexico, 
				82040 Mazatl\'an, M\'exico}
					\and
			Ana Carolina Ruiz-Fernandez\thanks{
				Unidad Acad\'emica de Mazatl\'an, 
				Instituto de Ciencias del Mar y Limnolog\'ia, 
				Universidad Nacional Aut\'onoma de Mexico, 
				82040 Mazatl\'an, M\'exico}

					\and
			J Andr\'es Christen\thanks{
				Centro de Investigaci\'on en Matem\'aticas (CIMAT),
				Jalisco s/n, Valenciana, 36023 Guanajuato, Gto, Mexico.
				email: \texttt{jac@cimat.mx}  }

			}
		\maketitle
	} \fi

	\if0\blind
	{
		\bigskip
		\bigskip
		\bigskip
		\begin{center}
			{\LARGE\bf \papertitle}
		\end{center}
		\medskip/
	} \fi

	\bigskip
\begin{abstract}
	The increasing interest in understanding anthropogenic impacts on the environment have led to a considerable number of studies focusing on sedimentary records for the last $\sim$ 100 - 200 years. 
Dating this period is often complicated by the poor resolution and large errors associated with radiocarbon (14C) ages, which is the most popular dating technique. 
To improve age-depth model resolution for the recent period, sediment dating with lead-210 ($^{210}$Pb) is widely used as it provides absolute and continuous dates for the last $\sim$ 100 – 150 years. 
The $^{210}$Pb dating method has traditionally relied on the Constant Rate of Supply (CRS, also known as Constant Flux - CF) model which uses the radioactive decay equation as an age-depth relationship resulting in a restrictive model to approximate dates. In this work, we compare the classical approach to $^{210}$Pb dating (CRS) and its Bayesian alternative (\textit{Plum}). 
To do so, we created simulated $^{210}$Pb profiles following three different sedimentation processes, complying with the assumptions imposed by the CRS model, and analysed them using both approaches. 
Results indicate that the CRS model does not capture the true values even with a high dating resolution for the sediment, nor improves does its accuracy improve as more information is available. 
On the other hand, the Bayesian alternative (\textit{Plum}) provides consistently more accurate results even with few samples, and its accuracy and precision constantly improves as more information is available.
\end{abstract}
	\noindent%
	\vfill
	\newpage
	\spacingset{1.45} 

\section{Introduction}

Lead-210 ($^{210}$Pb) is a radionuclide, part of the $^{238}$U decay chain, which forms naturally in the atmosphere as well as in sediments.
This isotope, with a half-life of 22.23$\pm$0.12 years, is commonly used to date recent recently accumulated sediments ($<150$years). 
In recent decades, increasing number of palaeoecological and pollution studies have focused on these recent sediments \citep[e.g.,][]{Courtney2019} in order to evaluate human impacts on the environment.
Unlike to other dating techniques such as $^{14}$C (radiocarbon dating), dating single sediment layers is not possible from a single measurement of $^{210}$Pb; it is only when a suitable portion of the excess-$^{210}$Pb (atmospheric $^{210}$Pb) decay curve is measured, representing the toral inventory of $^{210}$Pb from atmospheric deposition and runoff, and when certain assumptions about the sedimentation process are met that a chronology can be established.  
These studies strongly rely on the accuracy of their chronologies in order to correctly assign dates to chemical, biological and ecological changes.
That is, unlike other dating techniques, an analysis of a series (data set) of $^{210}$Pb measurements must be carried out in order to obtain meaningful dates.  Samples are taken along a core (e.g., lake, peatland, marine sediments) at different depths, from which $^{210}$Pb activity is measured.  
The whole series of $^{210}$Pb measurements need to be analysed in order to attempt to produce a coherent chronology, see \citet{Aquino2018}.

\begin{figure}[h!]
	\begin{centering}
		\includegraphics[width=.75\linewidth]{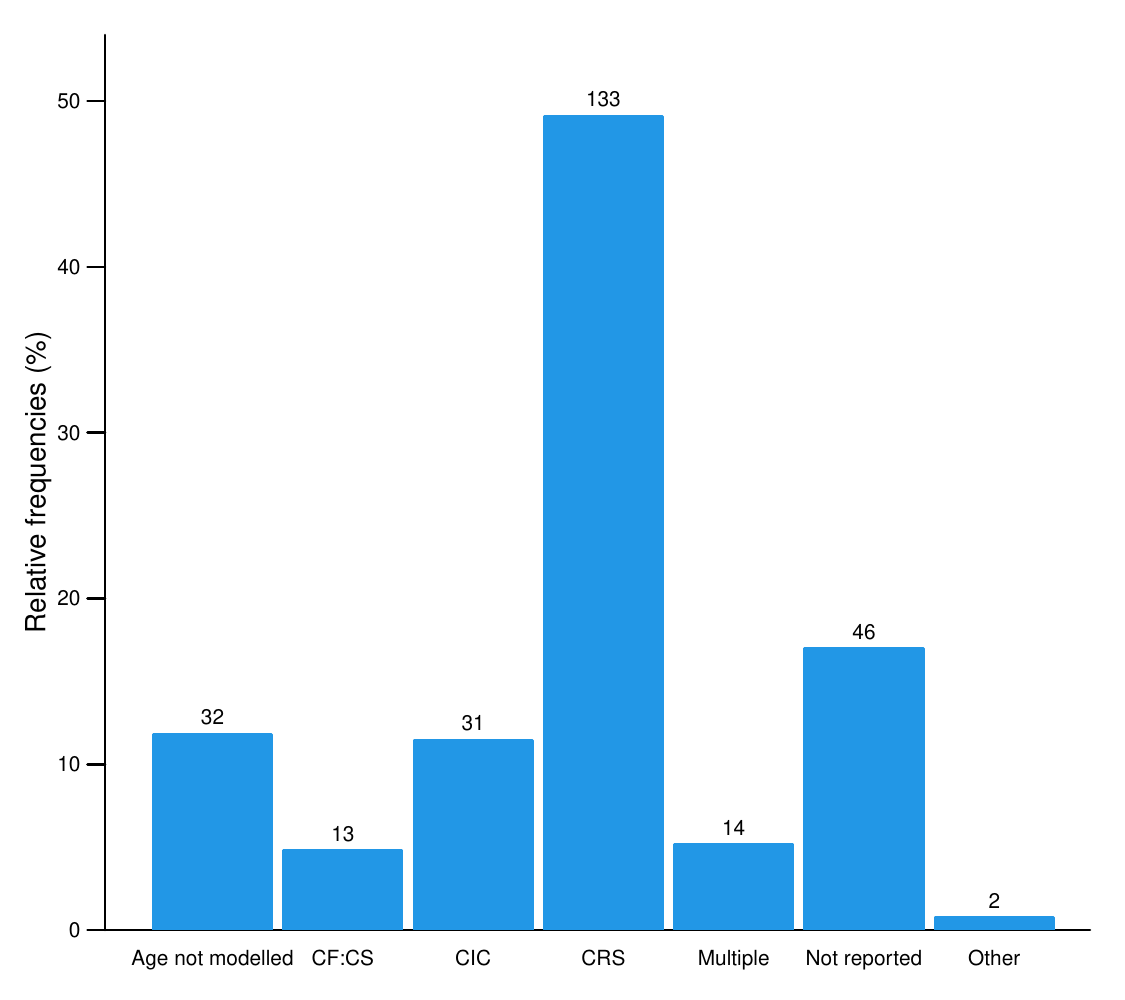}
		\caption{Frequency of $^{210}$Pb dating models used in papers between 1964 and 2017. Data gathered by \citet{Courtney2019} from a literature review of 271 papers. The models include CF:CS model \citep[The Constant Flux - Constant Sedimentation;][]{Robbins1978}, CIC (Constant Initial Concentration) \citep{Goldberg1963,Crozaz1964,Robbins1978} and CRS -  \citep[Constant Rate of Supply;][]{Appleby1978,Robbins1978}. }
		\label{fig:210models}
	\end{centering}
\end{figure}

A range of traditional data analyses, or so called ``models'', are available for dating recent sediments using $^{210}$Pb; e.g. the Constant Initial Concentration \citep[CIC,][]{Goldberg1963}, also known as Constant Activity \citep[CA,][]{Robbins1975}, the Constant Flux : Constant sedimentation \citep[CFCS,][]{Crozaz1964} and the Constant Rate of Supply  \citep[CRS,][]{Appleby1978,Robbins1978,Sanchez-Cabeza2012} also known as the Constant Flux model (CF). 
The CRS model is by far the most popular (see Figure \ref{fig:210models}) and has the most flexible assumptions. 
It assumes a constant supply of atmospheric $^{210}$Pb (also known as excess $^{210}$Pb) to the sediment and allows for changes in the sedimentation rate. 
In order to build a chronology, the CRS model uses a ratio between the complete ``inventory'' (the excess $^{210}Pb$ activity accumulated in the sediment column, between the surface and the equilibrium depth, where excess $^{210}$Pb  can no longer be found) and the remaining  inventory from depth $x$ to the previously defined equilibrium depth, ($t(x)=\frac{1}{\lambda}\log\left( \frac{A_0}{A_x}\right)$, where $A_0$ is the complete inventory and $\lambda$ the decay constant of the $^{210}$Pb $\approx 0.03118\pm 0.00017$ yr$^{-1}$).

Other, more restrictive, models such as CFCS and CIC also require the assumption of a constant flux of $^{210}$Pb, as well additional as other assumptions of the sedimentation process, as well as that of a constant supply of $^{210}$Pb.
The flexibility of the CRS model in terms of its assumptions, comes at the cost of needing to measure a sufficient portion of the inventory or to use of interpolation in order to properly estimate the complete inventory of $^{210}$Pb in the sediment. 

The CRS model has undergone several revisions in the last decade in order to improve its applicability and precision. 
There are two types of revisions to this model: (1) revisions to its uncertainty quantification \citep[eg. ][]{Binford1990,Appleby2001,Sanchez-Cabeza2014} and (2) to its application where extra information is available, such as external independent dating markers (e.g. $^{137}$Cs profiles), laminated sediments, tephras, contaminated layers (known sedimentary events) \citep[eg.][]{Appleby1998,Appleby2001,Appleby2008}. 

A recent inter-laboratory model comparison experiment \citep{Barsanti2020} presented concerning results.
Two measured $^{210}$Pb data sets were send to 14 laboratories around the world with varying degrees of expertise in the 210Pb dating method.
Each laboratory was asked to provide a chronology, given the same data. 
It is important to note that each laboratory applied their preferred model; in most cases the CRS model was calculated.
This experiment resulted in a wide range of chronologies, independently of the model used, providing different chronologies even when the same model and dataset was used.
The authors reinforced the need to use of independent time markers (independent dating sources) to validate and ``anchor" of the chronologies, as suggested previously by \citep{Smith2001}.  
This comparison experiment clearly and critically shows the impact that user decisions and applying expert adaptations/revisions have on the resulting chronologies.
In order to replicate and/or update any given chronology, such user decisions becomes extremely important.
In addition, raw data sets are also required; unfortunately, both the raw data sets and/or user's decisions are rarely reported.

Recently \citet{Aquino2018} presented an alternative to these classical models, by introducing \textit{Plum}, a Bayesian approach to $^{210}$Pb dating.
This model treats every data point as originating from a forward model that includes both the sedimentation process and the radioactive decay process.
\textit{Plum} also assumes a constant rate of supply to the sediment, similar to the CRS model (this assumption can be relaxed at the cost of computational power).Another important difference between the CRS and \textit{Plum} is that the latter incorporates the supported $^{210}$Pb, which naturally forms in the sediment and is normally threaded as a hindrance variable.
\textit{Plum} assumes that there exists an (unknown) age-depth function $t(x)$ that relates depth $x$ with calendar age $t(x)$. 
Conditional on $t(x)$, the following model is assumed for the measured $^{210}$Pb $y_i$ between depths $x_i - \delta$ to $x_i$
\begin{eqnarray}
y_i\mid P^S_i, \Phi_i, \bar{t}\sim \mathcal{N} \left(A^S_i+\frac{\Phi_i}{\lambda} \left( e^{-\lambda t(x_i-\delta)} - e^{-\lambda t(x_i)} \right), (\sigma_i\rho_i)^2 \right). 
\end{eqnarray}
Here $A_i^S$ is the supported $^{210}$Pb in the sample and $\Phi_i$ the supply of excess $^{210}$Pb to the sediment, the age-depth model $t(x)$ is based on a piece-wise linear model constrained by prior information on the sediment's accumulation rates  \citep{Blaauw2011}, see \cite{Aquino2018} for details.

This treatment of the data allows for a formal statistical inference on a well-defined model with specific parameters. 
In order to infer the parameters of the model, a Bayesian approach is used.
This differs from the CRS model, which does not provide a formal statistical inference.
The CRS model uses the decay equation to obtain an age-depth function, resulting in a more restrictive age-depth model. 
It only deals with the excess $^{210}$Pb, the estimated supported $^{210}$Pb having been previously removed before modelling.
\textit{Plum} has shown to provide accurate results with a realistic precision using different case scenarios \citep{Aquino2018,Aquino2020} - both in simulations as well as for real cores.
Under optimal dating conditions, \textit{Plum} and the CRS model have been shown to provide similar results \citep{Aquino2020}, with \textit{Plum} providing more realistic uncertainties, with minimal user interaction. 

\citet{Blaauw2018} presented a comparison between classical and Bayesian age-depth models construction, both for real and simulated $^{14}$C-dated cores.
They concluded that Bayesian age-depth models provide a more accurate result and more realistic uncertainties under a wide range of scenarios.  
The objective of the present study is to test whether the results obtained by \citet{Blaauw2018}, concerning the accuracy and precision of the Bayesian approach, are maintained in a more complex modelling situation, such as the construction of $^{210}$Pb-based age-depth models.
To do so, we compare $^{210}$Pb dates and uncertainties from the widely applied CRS model (by far the most popular age-depth model for $^{210}$Pb) against \textit{Plum} using simulated cores, i.e. sedimentation ``scenarios''.
We also aim to observe the learning process of each of the models and estimate the amount of information is needed to obtained a reasonable chronology for each model. 

Given that the CRS model has had several revisions, the choice of which can considerably affect model outputs as shown by \citet{Barsanti2020}, we decided to apply the original version of the equations provided by \citet{Appleby2001}, with its suggested error propagation calculation; we will call this version of the CRS the ``classical implementation of the CRS" (CI-CRS). 
hile we acknowledge that this implementation may be the less suitable in some particular cases and then expert knowledge can greatly improve the precision and accuracy of the model, but this will reduce the bias of any particular implementation has on our results.

The paper is organized as follows: second section sets the tools we use for the model comparison, describing the simulations of the three different scenarios.
Section 3 describes the comparison for both the overall chronologies and for single depths.
Section 4 shows the impact of expert revisions.
Lastly, Section 5 presents the conclusions and discussion of the results obtained in section 3.

\section{Experiment Setup: Simulations and Model Considerations}

	In order to observe the accuracy and precision of any chronology, a known true age-depth function is required.
\citet{Blaauw2018} presented a methodology for simulating radiocarbon dates and their uncertainties, while \citet{Aquino2018} presented an approach for simulating $^{210}$Pb data given an age-depth function $t(x)$.
It is important to note that these simulations follow the equations presented by \cite{Appleby1978, Robbins1978} guaranteeing that the CRS assumptions are met. 
By using the approach presented by \citet{Aquino2018} for simulating $^{210}$Pb data and the structure of uncertainty quantification presented by \citet{Blaauw2018}, reliable simulated $^{210}$Pb data can be obtained.

\subsection{Simulation Construction}\label{sec:SimConst}

\begin{figure}[!h]
 \centering
  \includegraphics[width=.95\linewidth]{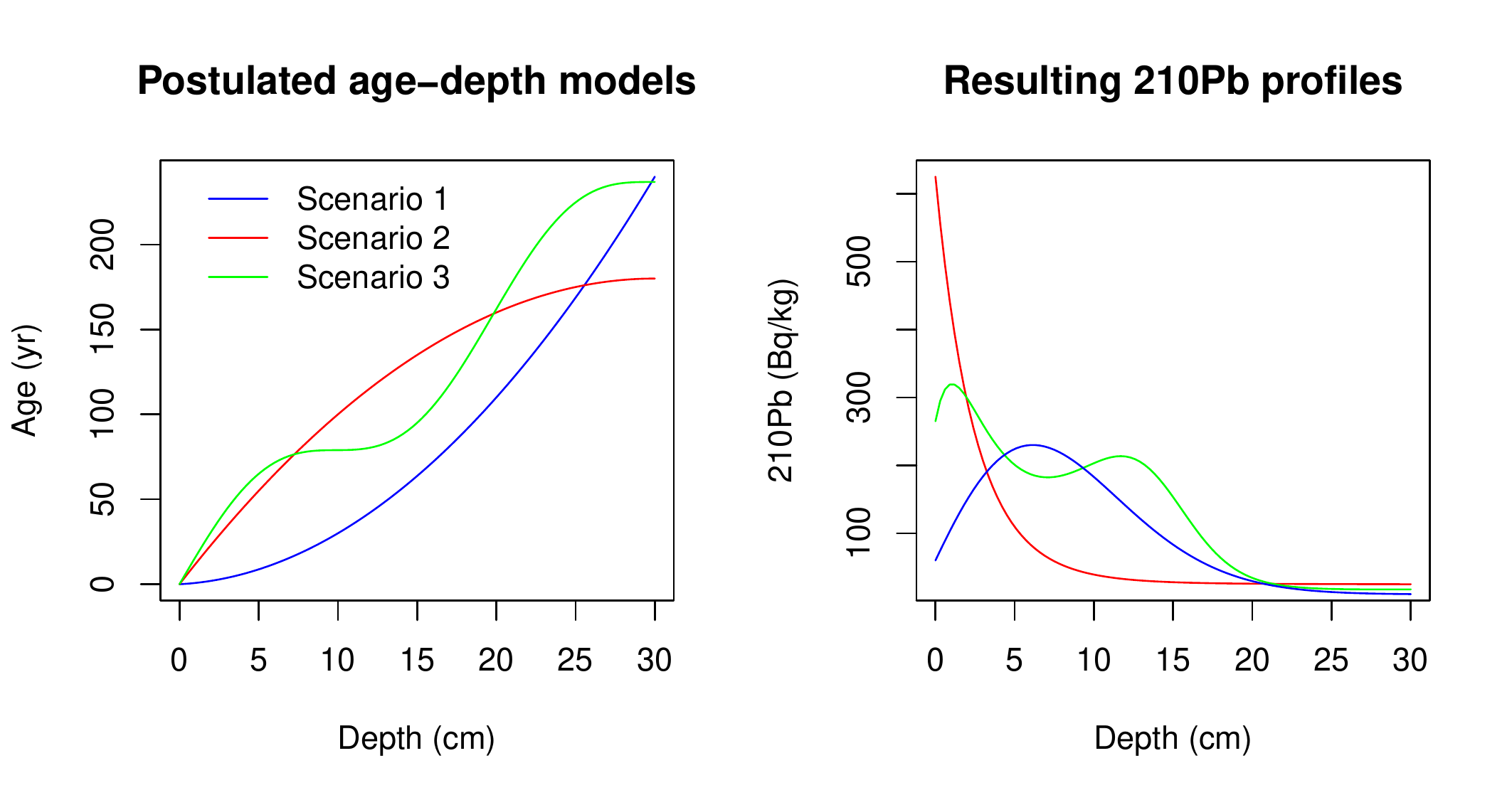}
	\caption{Simulated sedimentation scenarios with their corresponding $^{210}$Pb profiles. Left: Age-depth functions for the three different scenarios (Table \ref{tab:sim_param}). Right: Corresponding $^{210}$Pb activity profiles in relation to depth.}
  \label{fig:true_210}
\end{figure}

Three different scenarios (see Table \ref{tab:sim_param}) were chosen to simulate sedimentation processes, with their own age-depth functions and parameters. 
These scenarios were selected as they provide three key challenges for the models: Scenario 1 presents an age-depth function which is the result of increasing sedimentation and less compaction towards the present (surface); this is quite common for recent sediments. Scenario 2 presents a challenging core structure as the function has a drastic and rapid shift in sediment accumulation around depth 15 cm depth. Lastly Scenario 3 presents a cyclic and periodic change in accumulation rates. 
Using the age-depth functions and parameters defined in Table \ref{tab:sim_param}, we obtain the $^{210}$Pb activity, or concentration, at any given depth or interval, by integrating the age-depth curve for that interval.  
Although these concentrations may be interpreted as error-free measurements 
(see Figure \ref{fig:true_210}), we replicated the $^{210}$Pb activity uncertainty, following a similar methodology to \citet{Blaauw2018}.
This methodology was chosen as it introduces different sources of uncertainty related to different steps of the measurement process.
Other uncertainty quantification methodologies could be used, but as long as the same methodology and uncertainty is provided to both models the comparison remains valid.
\begin{table}[!h]
	\centering
	\begin{tabular}{l|ccc}
Label    	& 	Age-depth		&	$ \Phi$		& Supported $^{210}$Pb  \\
		&	function $t(x)$		&	($\frac{Bq}{m^2yr }$)	& ($\frac{Bq}{kg}$) 	\\ \hline
Scenario 1 	&	$\frac{x^2}{4} + \frac{x}{2}$	&	100	& 10	\\
Scenario 2 	&	$12x -.2x^2$			&	50	& 25	\\
Scenario 3 	&	$8x+25\sin(\frac{x}{\pi})$	&	500 	& 15		
	\end{tabular}
	\label{tab:sim_param}
	\caption{Simulated age-depth function and parameters used in each scenario}
 \end{table}

	Let $C_{\hat{x}}$ be the true $^{210}$Pb concentration in the interval $\hat{x}=[ x-\delta, x)$, given the age-depth function $t(x)$ and parameters $\Phi$ and $A^S$ in each scenario. 
To simulate disturbances in the material, we can introduce scatter centred around the true value, $\theta \sim \mathcal{N}\left(C_{\hat{x}},y^2_{scat}\right)$, where $x^2_{scat}$ is the amount of scatter for this variable (in this case $y^2_{scat}=10$). 
Now, to replicate outliers, a shift from the true value ($x_{shift}$) is defined, which occurs with a probability $p_{out}$. This results in a new variable $\theta'$ which is defined as
\begin{align}
	\theta' = \begin{cases}
			\mathcal{U}(\theta - x_{shift},\theta + x_{shift}), &  p_{out} \\
			\theta, & 1-p_{out}
		\end{cases}.
\end{align}

	Finally, to simulate the uncertainty provided by the laboratory, we can define the simulated measurements as  $y(\theta')\sim\mathcal{N}\left(\theta',\sigma_R^2\right)$, where $\sigma_R$ is the standard deviation reported by the laboratory. 
$\sigma_R$ is defined as $\sigma_R= \max \left(\sigma_{min}, \mu(\theta')~\varepsilon~y_{scat}~\right)$, where $\sigma_{min}$ is the minimum standard deviation assigned to a measurement. This variable differs between laboratories,we use a default value of $1~ Bq/kg$. 
Finally, $\varepsilon$ is the analytical uncertainty (default .01) and $y_{scat}$ an error multiplier (default 1.5).
The default parameters were set in accordance with \citet{Blaauw2018}.

For this this study we created a data set for each of the three simulation by integrating in intervals of $\delta =$1 cm, for depths from  0 to 30 cm where radioactive equilibrium was guaranteed \citep{Aquino2018}.
The complete simulated $^{210}$Pb data sets can be found in the Supplementary Material \ref{sec:supp_mat}.

\subsection{Model Considerations}
In order to create a comparison with minimal user interaction, each model was run automatically.
In the case of \textit{Plum}, default settings were used in order to minimize user interaction.
As the CI-CRS model assumes that background (supported) $^{210}$Pb has been reached, in order to reduce user manipulation, we decided to fix the last sample (30 cm depth) for every case.
This step not only guarantees the consistent application of the CI-CRS model, it also provides the model with a single bottom-most depth to be removed as it is common practice when using the CI-CRS model.
Because of this, \textit{Plum}'s resulting chronology will always reaches 30 cm, as by default 1 cm sections are used for every simulation.
Conversely, as CI-CRS model only models the excess $^{210}$Pb (the total $^{210}$Pb minus the supported $^{210}$Pb), when certain excess activities at depth fall below zero, the chronology will only be calculated up to that depth.
\textit{Plum} deals with this variable supported $^{210}$Pb variable automatically, as part of the inference.
In order to provide the best possible estimate for this variable a constant level of supported $^{210}$Pb was assumed for both models.
For the CI-CRS model, the mean of the supported $^{210}$Pb measurements was calculated and then subtracted from the total $^{210}$Pb to obtain the excess $^{210}$Pb, as it is common practice when using the CI-CRS model.

In order to provide an objective comparison, the offset of the true age-depth model (in yr), length of the 95\% intervals (in yr) and normalized accuracy were calculated (the normalized offset indicates the distance of modelled ages from the true value given the model's own uncertainty). 
The main discussion will revolve around the normalized offset as it provide an intuitive measure of the accuracy a model by taking into account the levels of uncertainty provided by each model. 

\section{Model Comparison}

\begin{figure}[!]
	\centering
	\includegraphics[width=\linewidth]{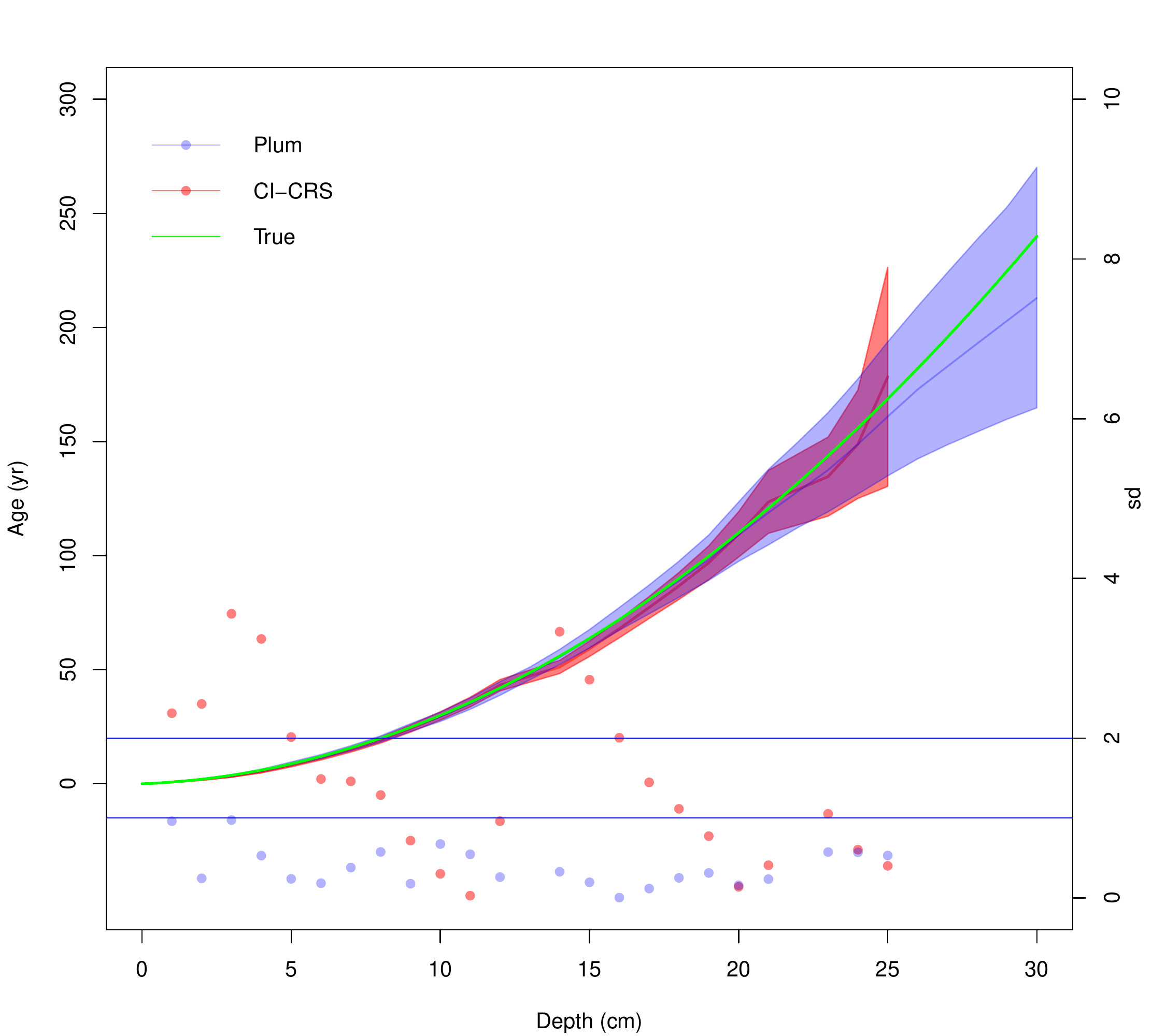}
		\caption{Comparison between \textit{Plum} and the CI-CRS model against the true age-depth model using 95\% of the information percentage (using 1-cm samples at depths at every depth except 13 and 19 cm). Lines show the age estimates with the 95\% credible intervals (\textit{Plum}) and the 95\% confidence interval (CI-CRS). Dots show the normalized offset, i.e. the distance between the inferred age and the true age in relation to the standard error (the standard deviation in the case of the CI-CRS and the length of the confidence interval divided by 4 in the case of \textit{Plum}). The vertical right axis shows how many standard deviations is each model from the true age.  }
		\label{fig:comparison1r}
\end{figure}

To allow for a reasonable comparison between models, and to evaluate the effect that different amount of information may have on the accuracy and precision of $^{210}$Pb models, we used our three simulated data sets (see Supplementary Material \ref{sec:supp_mat}). 
For these simulated cores, samples were randomly generated, given a percentage of information (e.g. for a 20\% information a dataset with 6 random 1-cm samples -of a possible total 30 1-cm samples- is created) in order to create a sub-dataset, which was then used to create a chronology:
100 of these sub-datasets were created for information percentages from 10\% to 95\% at 5\% intervals (i.e., 10\%, 15\%, 20\%,...,95\%). 
The complete dataset was also used (i.e 100\% percentage of information sample).
Once a dataset was created, both the CRS model and \textit{Plum} were applied.  
Both sets of outputs were then compared against the true known age value, see Figure \ref{fig:comparison1r}.

Figure \ref{fig:comparison1r} shows a single ``snapshot" an example of the comparison between the $^{210}$Pb models against the true value. 
As we are dealing with a total of $n =$ 5333 simulations, in order to evaluate the overall precision and accuracy of both models, we decided to calculate the mean offset to the true age-depth model (in yr), the mean of length of the 95\% intervals (in yr), as well as the mean normalized accuracy indicating the distance of modelled ages from the true value given the model's own uncertainty at each depth.  

\begin{figure}[!]
 \centering
  \includegraphics[width=.75\linewidth]{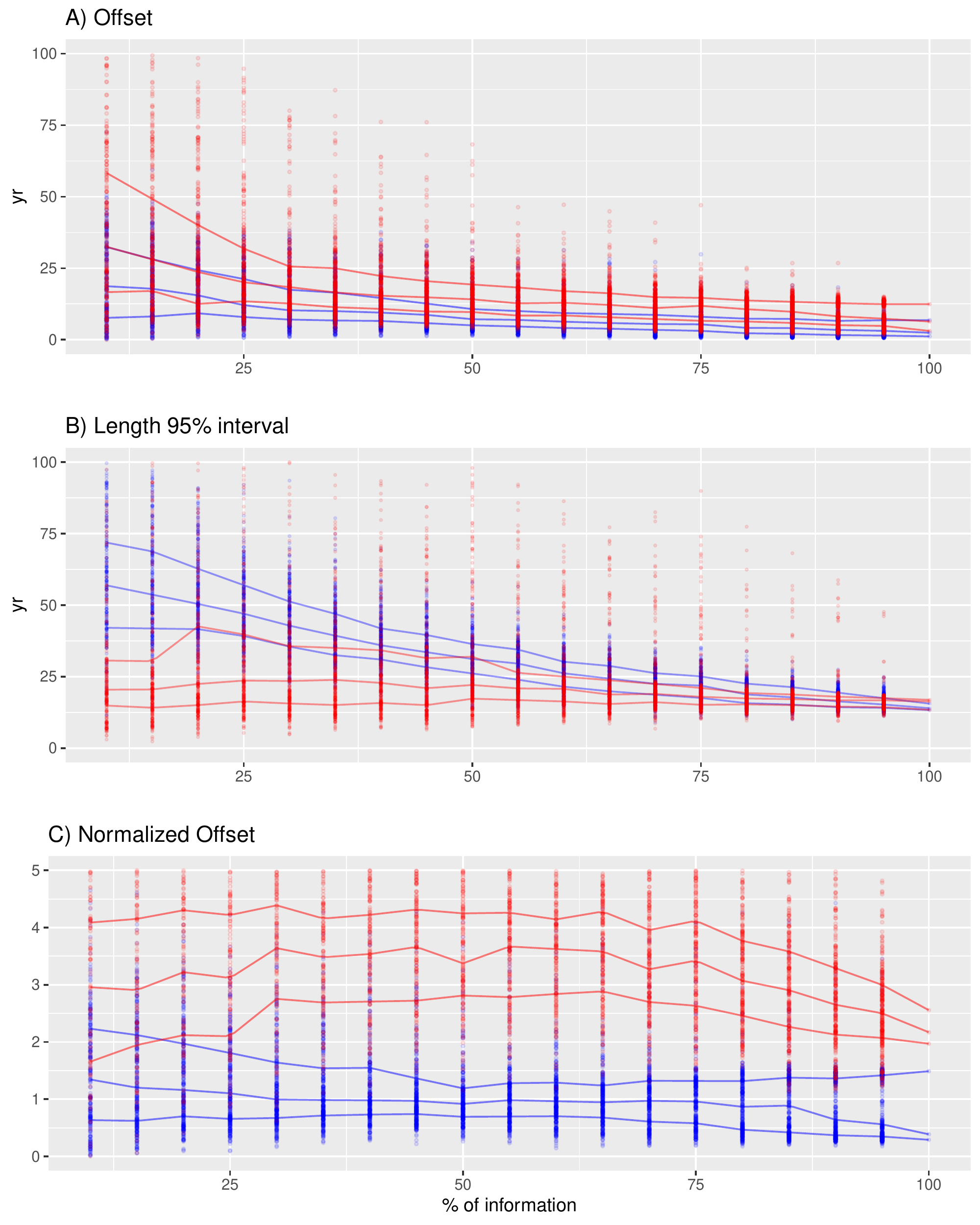}
	\caption{Top panel A) shows the offset between the modelled and true age of the CI-CRS (red) and \textit{Plum} (blue). This panel shows how \textit{Plum} provides a small offset in almost every scenario with both models improving their offset as more information is available. Middle panel B) shows the 95\% confidence intervals. It is clear, from this panel, than the uncertainty provided by \textit{Plum} is a lot bigger for low percentage of information and it constantly improves as more data is available, whereas the length of the intervals provided by the CI-CRS appear to stay constant regardless of the available information. Bottom panel C) shows the normalized offsets, presenting the distance between the modelled age and the true age normalized divided by the standard deviation (in the case of \textit{Plum}, the length of the 95\% interval divided by 4). This panel presents a worrying situation where the CI-CRS model's calculated standard deviation (on average) is incapable of capturing the true age. On the other hand, \textit{Plum}'s credible intervals almost always capture the true age even when little information is available.}
  \label{fig:accpre}
\end{figure}

Figure \ref{fig:accpre} show results similar to those presented by \citet{Blaauw2018}. 
The classical model (CI-CRS) at first appears to provide a similar results (similar offsets) to the Bayesian alternative (\textit{Plum}), but at higher estimated precision, if we only consider at the length of the 95\% interval. 
It is important to note that the CI-CRS model's offset improves as more information is available.
However, if we do not consider both the effects of both the offset and length of the interval together, the results are not favourable to the CI-CRS. 
To have a more realistic representation of how the models capture the true age-depth models relationship, we should observe the normalized offset. 
This variable shows the degree to which the average models contain the truth within their uncertainty intervals (normalized to one standard deviation). 
Any model with a normalized offset larger than two (two standard deviations) is incapable of capturing the true ages within its uncertainty intervals.  
This means that, while the CI-CRS estimates smaller uncertainties and its ages improve as more data is available, it does so at the cost of its accuracy and the improvements are not sufficient to capture the true age.
It also appears that the length of the 95\% interval and offset are not affected by how much information is provided to the CRS model. 

On the other hand, \textit{Plum} seems to provide increasingly accurate results as more information is added to the model.
This again coincides with the results outlined by \citet{Blaauw2018}. 
When we observe the regular offset (not normalized), we find that \textit{Plum} provides a smaller offset in comparison to the CI-CRS model; this, in combination with slightly larger (more realistic) modelled uncertainties, results in more consistently accurate age-depth models that are capable of capturing the true values within their uncertainty intervals. 
This result supports the claim that \textit{Plum} provides more realistic uncertainties compared those obtained by the CI-CRS. 
Another important statistic to take into account is that 87.86\% (4686/5333) of \textit{Plum}'s runs remain within the 2 standard deviations, opposed to 7.48\% (399/5333) for the CI-CRS model. Furthermore, only 0.54\% (29/5333) of the CI-CRS model runs remain under the 1 standard deviation, which is the most commonly reported interval when reporting CI-CRS results.
We can also observe a clear structure in the way that \textit{Plum} increases its accuracy and precision to obtain a better chronology as more information is available, whereas the CI-CRS model does not appears to improve its ability to capture the true value from additional data. 

These results are valid for the overall chronology (the mean offset, interval and normalized offset of the overall chronology). 
In order to evaluate whether certain models are better predicting ages at certain section of the sediment cores, we have to look at the normalized offset of every depth.


\begin{figure}[!]
	\begin{centering}
		\includegraphics[width=\linewidth]{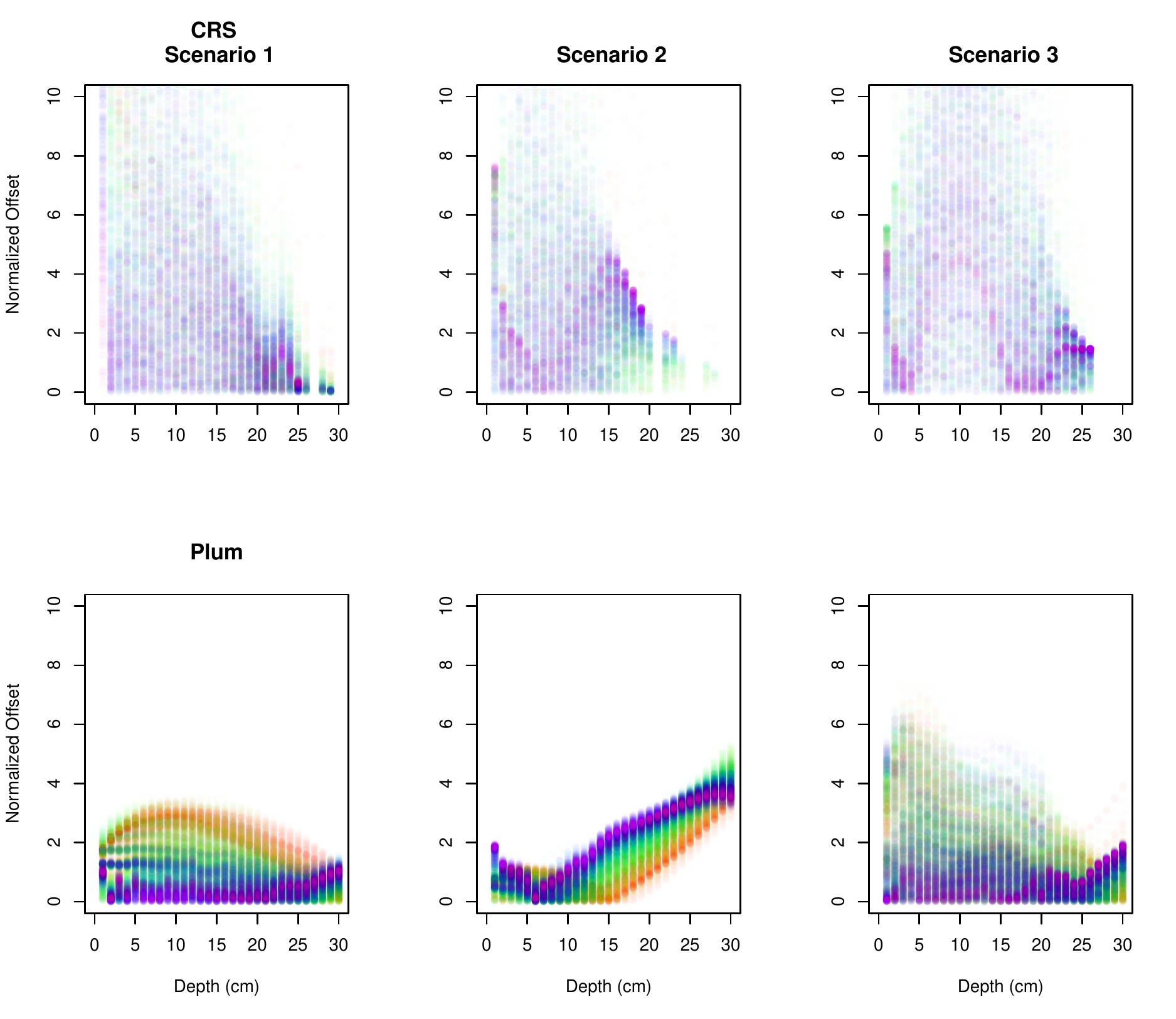}
		\caption{Normalized offset of every sampling sample at every depth for the three simulated scenarios - CI-CRS age estimates at samples depths and \textit{Plum}'s age estimates at 1 cm intervals. Dots go from lowest information percentage samples (few dated depths; red) to high percentage samples (nearly completely dated cores; purple). The CI-CRS's normalized offset shows no learning pattern at any particular depth regardless of the available information. This means that the model can provide a reasonable chronology with low levels of information or a very inaccurate age estimate with high levels of information at any given depth resulting in a distrustful age-depth model. On the other hand, \textit{Plum} demonstrares a systematic improvement in its age estimates as more data is available. This results assures that a Bayesian approach would consistently provide more reliable results.     }
		\label{fig:depths}
	\end{centering}
\end{figure}

Figure \ref{fig:depths} shows the normalized accuracy of every simulation according to depth for both models.
\textit{Plum} shows a clear learning structure which depends on the information available to the model.
The information percentage appears to be irrelevant to the normalized accuracy of the CI-CRS model, contrary to the results obtained by \textit{Plum}.
It is important to note that the inaccuracies of the CI-CRS model are not exclusive to any particular sections of the chronology; this is most likely driven by the small uncertainties estimated by the CI-CRS model.See below for a discussion of how \textit{Plum} behaved in sedimentation simulation 2.

\section{Improvements to the CI-CRS}

Since the late 1970's, when the CRS method was first introduced \citep{Appleby1978,Robbins1978}, the CRS has undergone several improvements. 
Some of these improvements rely on independent dates, other isotopes or techniques, and/or require user manipulation to ``force" the method to agree with these independent dates.
One recent improvement, which does not require user manipulation and/or independent dates, is the comprehensive explanation, with expert notes, on the practical used of the CRS model by \citet{Sanchez-Cabeza2012}. 
The same authors presented an improvement to the uncertainty quantification of the age estimates by using the Monte Carlo method \citep{Sanchez-Cabeza2014}
and release a publicly available Excel spreadsheet, which facilitate the calculation of their age estimates and Monte Carlo uncertainties. 
\citet{Barsanti2020} showed that there exist several modifications and improvements to the CRS, and that the choice of modifications can generate a range of age-depth models.
Considering that this research focuses on the methods with minimal user manipulation, and given that these modifications are laboratory-specific and not made publicly available, an R implementation of the improved CRS by \citet{Sanchez-Cabeza2014}, here labelled as revised CRS (R-CRS), was used to calculate a chronology of a particular, randomly selected, subsample with 95\% of information and then compared against \textit{Plum} and CI-CRS.
The goal of this experiment is to quantify the improvement than a particular modification may have on the age estimates provided by the CRS methodology.

\begin{figure}[h!]
 \centering
  \includegraphics[width=.95\linewidth]{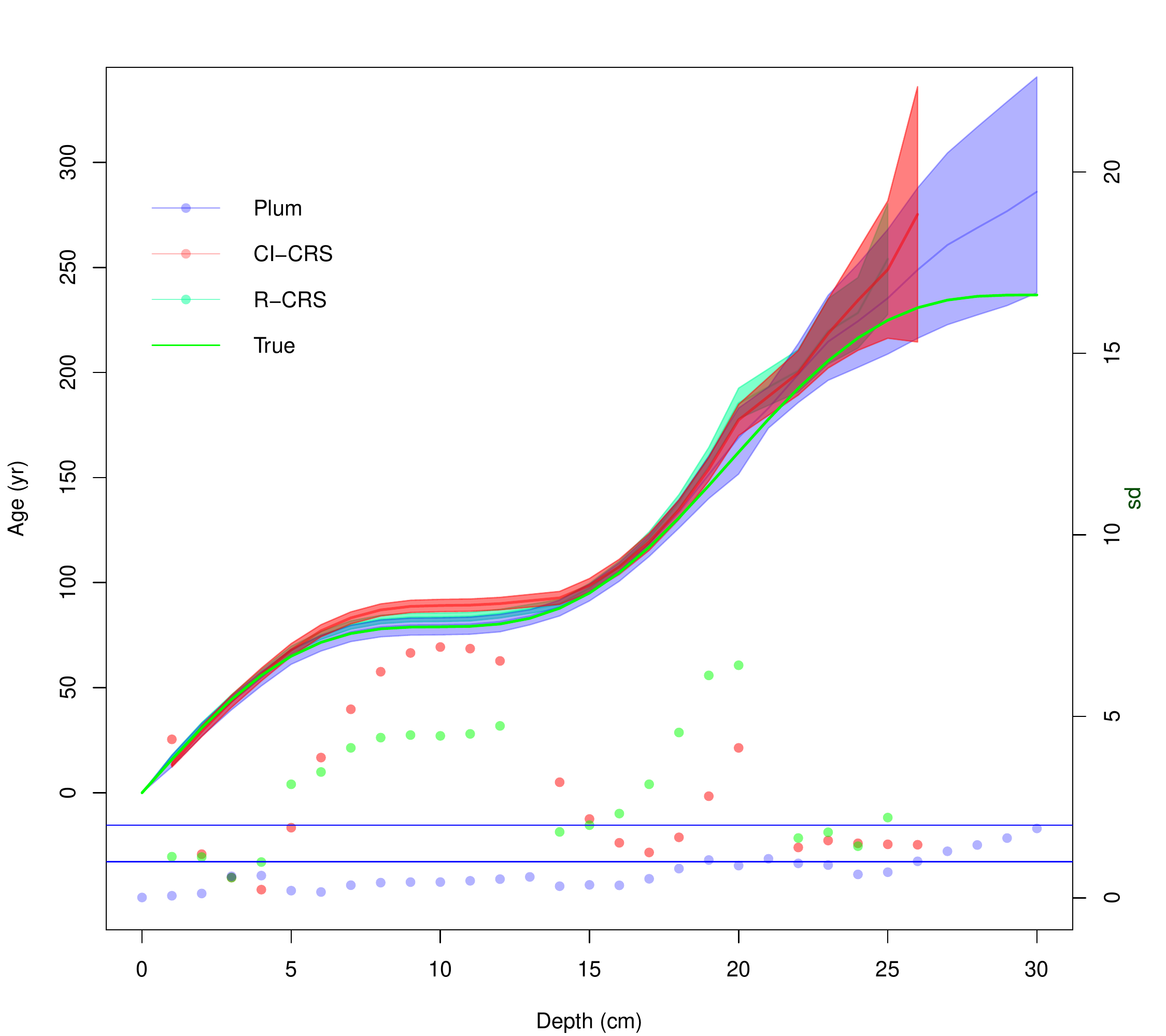}
	\caption{Comparison between the CI-CRS, R-CRS and \textit{Plum}.} 
  \label{fig:95compa}
\end{figure}

Figure \ref{fig:95compa} shows the resulting chronologies for the three methodologies.
From this figure it is clear that both versions of the CRS model (CI-CRS and R-CRS), with a good percentage of information, are capable of replicating the cyclic changes in accumulation that this scenario presents.
It is also important to note that the offset of the R-CRS is improved throughout the whole chronology. 
However, the smaller uncertainties provided by the Monte Carlo method in \citet{Sanchez-Cabeza2014} appear to be the reason that the normalized offset, of the R-CRS, is larger in some parts of the chronology, when compared to the CI-CRS. 

The discussion on how large the standard deviation has to be in order to properly represent the uncertainty of the CRS model goes beyond the scope of this research.
Nevertheless, it is important to point out that the realistic uncertainties provided by \textit{Plum}, are the result of a proper statistical inference where a likelihood is defined.
In the case of the two version of the CRS model used in this research (CI-CRS and R-CRS) a proper likelihood is never defined and the uncertainties are calculated as a separate part of the model, either by error propagation or the Monte Carlo method.
This could be the reason that the uncertainty quantification appears to be unrepresentative.

\section{Discussion and Conclusions}

This research focuses on exploring the uncertainty and precision of the most commonly used $^{210}$Pb dating methods (CI-CRS and \textit{Plum}).
By using different scenarios, three different simulations were created.
These simulations were then sub-sampled at different percentages of information in order to observe the effects that different sample sizes have on the resulting chronology. 
This experiment provided an objective comparison of the accuracy and precision of both methods.

The experiment was measured in two different level.
For the first level, the overall accuracy and precision of the method were evaluated.
The mean of the offset, length of the 95\% confidence and credible intervals, as well as the normalized offset were measured.
The second level focused on the ability of each model to capture the true value in their credible/confidence interval, and the normalized offset of every scenario per depth was calculated. 
These two comparisons provided a good picture of the difference in precision and accuracy between these methods.

From the overall accuracy (see Figure \ref{fig:accpre}) it is clear that both the CI-CRS model and \textit{Plum} reduce their offset as more data becomes available, with the Bayesian method providing, on average, a smaller offset regardless of the sample size. 
On the subject of precision, the Bayesian method is providing much larger uncertainties when small sample sizes are used. 
It is only with 60\%, or more, of the information that the length of the intervals becomes comparable. 
This is a consequence of the linear interpolation, between data points, used by the CRS method, in contraste to the Bayesian approach (\textit{Plum}) using a proper statistical inference.  
As has been previously discussed by \citet{Aquino2020}, the larger uncertainties provided by \textit{Plum} are more realistic, and this experiment confirms the latter.
Further evidence that these uncertainties are more sensible is that the length of the credible intervals becomes smaller as more data becomes available. 
On the other hand, the length of the confidence intervals provided by the classical model (CI-CRS) remain almost constant at any sample size.
Lastly, the normalized offset, which shows the capability of the model to capture the true values within their intervals, shows that the classical model (CI-CRS) on average is incapable of capturing the true values within its 95\% confidence interval. 
These results are especially worrisome considering that the $^{210}$Pb dating community rarely report 95\% confidence intervals and instead tend to use only 65\% confidence intervals (one standard deviation intervals) are reported.
On the other hand, \textit{Plum}'s normalized offsets always remain $\leq 2$, therefore guaranteeing that on average the true value is capture within its 95\% credible intervals, even with small sample sizes.
\textit{Plum}'s normalized offsets are constantly improving and reaching stability with 50\%, or more, of information percentage.
These experiments show that the Bayesian method, on average, provides more reliable results.

Because the normalized offset shows the capability of capturing the true value within its intervals, this variable can be used to conclude if any given method is better at estimating certain time period.
Figure \ref{fig:depths} presents the performance of both the CI-CRS model and \textit{Plum} for every simulated scenario.
It appears that, the normalized offset of many of the CI-CRS chronologies are $> 2$ throughout the whole chronology, meaning that the model does not have a period of time for which it is more precise. 
Moreover, the CI-CRS does not exhibits a clear learning pattern, where the normalized offset appears to be indifferent to the amount of information available.
It appears that even high levels of information percentage provide normalized offsets $> 2$, in some cases closer to 4 for scenarios 2 and 3.
\textit{Plum} on the other hand, shows a structure where more data is reflected in improved models in scenarios 1 and 3.
It is only at low levels of information where \textit{Plum}'s normalized offset is $>2$.
Scenario 2, on the other hand, presents a case where \textit{Plum} is both incapable of capturing the true value, for depths deeper than 15 cm, and it appears that as more data becomes available the model provides worse results. 
This may be of concern if we do not recognized that this scenario is unrealistic as it presents an extreme change in the accumulation around 15 cm, which coincides with the depth at which the normalized offset becomes $>2$.
It is also important to acknowledge that this experiment was performed using default settings.  
In a real-world scenario the user typically has some prior knowledge of the sedimentation process, about the site of interest, which could be incorporated as prior information to the model to improve the resulting chronology.

The results obtained by this experiment appear to persist even in the case of the revised version of the CRS model (R-CRS).
The R-CRS model appears to improve the offset but this improvement appears to be nullified by the smaller uncertainties presented by \citet{Sanchez-Cabeza2014}.
The question of which version of the CRS provides the best result is beyond the scope of this research and is dependent on expert application of the model.
Nevertheless, it is important to note that that the offset, related to the CI-CRS and R-CRS, are reasonably small at certain sections of the sediment, but the uncertainty quantification of both methods is overly optimistic.  

In conclusion, the use of the Bayesian age-depth models is preferred for the consistent construction of sediment chronologies, not only on radiocarbon-based chronologies as presented by \citet{Blaauw2018} but also in the more complex case of $^{210}$Pb as demonstrated by this research.
While the classical approach provides reasonable results, regarding the offset, unfortunately the uncertainty quantification in these methods needs improvements as they do not rely on a proper statistical structure. 
In a real-world scenario, it is impossible to measure the true offset of a method and therefore a proper uncertainty quantification becomes extremely important.
These results support the recommendations presented by \citet{Smith2001,Barsanti2020} where the CRS method, or any dating methodology, should be validated using independent dating methods markers. 

Lastly, it is important to highlight the benefits of the Bayesian methods.
From both \citet{Blaauw2018} and the present work, it is shown that Bayesian methods constantly improve as more data are added, the uncertainty associated to the method is realistic and coherent with the amount of information available. This leads to chronologies that are capable of capturing the true age in their credible intervals. 
The ability to capture the true value in the credible intervals becomes important when the problem is associated with decision making processes, as it provides a more realistic picture of the available knowledge of the process. 
Given that $^{210}$Pb dating is now widely-used in pollution, environmental and climate change studies, which potentially have a high impact on both policy making and public perception, realistic age estimates and uncertainties become extremely important.

\section{Acknowledgments}

The authors are partially founded by CONACYT CB-2016-01-284451 and COVID19 312772 grants and a RDCOMM grant.
The corresponding author is founded by CONACYT through the postdoctoral residence program with CVU  489201.

\bibliographystyle{apalike}
\bibliography{bibliography.bib}
\newpage

\section{Supplementary Material}
\label{sec:supp_mat}
Data for each simulation and code used is hosted at: https://github.com/maquinolopez/Paper\_Simulations
\newpage
\begin{table}[h]
	\begin{tabular}{c|cllllll}
		Label    & Depth & Density  & 210Pb & sd(210Pb) & Thickness& 226Ra  & sd(226Ra) \\
		& (cm) &($g/cm^3$) &(Bq/kg)& & (cm) & (Bq/kg)&\\
		\hline 
		Sim01-01 & 1          & 0.10009                         & 63.50103      & 2.85755   & 1              & 23.8045       & 1.125     \\
		Sim01-02 & 2          & 0.10064                         & 80.08738      & 3.60393   & 1              & 23.2924       & 1.125     \\
		Sim01-03 & 3          & 0.10173                         & 98.32806      & 4.42476   & 1              & 23.434        & 1.125     \\
		Sim01-04 & 4          & 0.10334                         & 125.45705     & 5.64557   & 1              & 26.0873       & 1.125     \\
		Sim01-05 & 5          & 0.10547                         & 141.27971     & 6.35759   & 1              & 22.8041       & 1.125     \\
		Sim01-06 & 6          & 0.10809                         & 130.27571     & 5.86241   & 1              & 23.4333       & 1.125     \\
		Sim01-07 & 7          & 0.11116                         & 134.04051     & 6.03182   & 1              & 25.6156       & 1.125     \\
		Sim01-08 & 8          & 0.11466                         & 129.69245     & 5.83616   & 1              & 26.1371       & 1.125     \\
		Sim01-09 & 9          & 0.11855                         & 134.93655     & 6.07214   & 1              & 25.4813       & 1.125     \\
		Sim01-10 & 10         & 0.12278                         & 109.39886     & 4.92295   & 1              & 25.8877       & 1.125     \\
		Sim01-11 & 11         & 0.12731                         & 110.68133     & 4.98066   & 1              & 24.4414       & 1.125     \\
		Sim01-12 & 12         & 0.13209                         & 102.38094     & 4.60714   & 1              & 24.9053       & 1.125     \\
		Sim01-13 & 13         & 0.13706                         & 75.80895      & 3.4114    & 1              & 22.9151       & 1.125     \\
		Sim01-14 & 14         & 0.14218                         & 77.60406      & 3.49218   & 1              & 24.4808       & 1.125     \\
		Sim01-15 & 15         & 0.14738                         & 68.4401       & 3.0798    & 1              & 24.9343       & 1.125     \\
		Sim01-16 & 16         & 0.15262                         & 60.72037      & 2.73242   & 1              & 25.2659       & 1.125     \\
		Sim01-17 & 17         & 0.15782                         & 50.28147      & 2.26267   & 1              & 22.961        & 1.125     \\
		Sim01-18 & 18         & 0.16294                         & 44.24641      & 1.99109   & 1              & 22.9139       & 1.125     \\
		Sim01-19 & 19         & 0.16791                         & 39.85997      & 1.7937    & 1              & 28.3774       & 1.125     \\
		Sim01-20 & 20         & 0.17269                         & 38.40823      & 1.72837   & 1              & 23.5379       & 1.125     \\
		Sim01-21 & 21         & 0.17722                         & 32.75922      & 1.47416   & 1              & 25.4363       & 1.125     \\
		Sim01-22 & 22         & 0.18145                         & 28.02545      & 1.26115   & 1              & 24.8995       & 1.125     \\
		Sim01-23 & 23         & 0.18534                         & 27.8749       & 1.25437   & 1              & 22.6783       & 1.125     \\
		Sim01-24 & 24         & 0.18884                         & 30.74797      & 1.38366   & 1              & 24.8575       & 1.125     \\
		Sim01-25 & 25         & 0.19191                         & 28.36187      & 1.27628   & 1              & 24.8724       & 1.125     \\
		Sim01-26 & 26         & 0.19453                         & 27.24535      & 1.22604   & 1              & 24.3778       & 1.125     \\
		Sim01-27 & 27         & 0.19666                         & 23.59236      & 1.06166   & 1              & 24.7209       & 1.125     \\
		Sim01-28 & 28         & 0.19827                         & 25.74855      & 1.15868   & 1              & 24.6615       & 1.125     \\
		Sim01-29 & 29         & 0.19936                         & 25.05368      & 1.12742   & 1              & 24.7199       & 1.125     \\
		Sim01-30 & 30         & 0.19991                         & 25.0065       & 1.12529   & 1              & 24.4937       & 1.125    
	\end{tabular}
\end{table}

\begin{table}[h]
	\begin{tabular}{c|cllllll}
		Label    & Depth& Density& 210Pb & sd(210Pb) & Thickness & 226Ra & sd(226Ra) \\
				& (cm) &($g/cm^3$) &(Bq/kg)& & (cm) & (Bq/kg)&\\
		\hline 
		Sim02-01 & 1          & 0.1001                          & 909.3928      & 40.9227   & 1              & 8.9761        & 0.45      \\
		Sim02-02 & 2          & 0.1006                          & 683.9989      & 30.7799   & 1              & 10.0607       & 0.45      \\
		Sim02-03 & 3          & 0.1017                          & 453.0503      & 20.3873   & 1              & 9.8701        & 0.45      \\
		Sim02-04 & 4          & 0.1033                          & 310.7897      & 13.9855   & 1              & 10.37         & 0.45      \\
		Sim02-05 & 5          & 0.1055                          & 218.0058      & 9.8103    & 1              & 10.0418       & 0.45      \\
		Sim02-06 & 6          & 0.1081                          & 158.6974      & 7.1414    & 1              & 10.104        & 0.45      \\
		Sim02-07 & 7          & 0.1112                          & 113.9062      & 5.1258    & 1              & 10.2049       & 0.45      \\
		Sim02-08 & 8          & 0.1147                          & 75.5493       & 3.3997    & 1              & 9.334         & 0.45      \\
		Sim02-09 & 9          & 0.1185                          & 56.6252       & 2.5481    & 1              & 10.5145       & 0.45      \\
		Sim02-10 & 10         & 0.1228                          & 44.1595       & 1.9872    & 1              & 9.8677        & 0.45      \\
		Sim02-11 & 11         & 0.1273                          & 34.7448       & 1.5635    & 1              & 9.7694        & 0.45      \\
		Sim02-12 & 12         & 0.1321                          & 25.384        & 1.1423    & 1              & 10.5134       & 0.45      \\
		Sim02-13 & 13         & 0.1371                          & 24.0007       & 1.08      & 1              & 10.4589       & 0.45      \\
		Sim02-14 & 14         & 0.1422                          & 21.3643       & 1         & 1              & 9.9504        & 0.45      \\
		Sim02-15 & 15         & 0.1474                          & 17.7932       & 1         & 1              & 10.5135       & 0.45      \\
		Sim02-16 & 16         & 0.1526                          & 15.0416       & 1         & 1              & 10.3362       & 0.45      \\
		Sim02-17 & 17         & 0.1578                          & 14.2937       & 1         & 1              & 10.5131       & 0.45      \\
		Sim02-18 & 18         & 0.1629                          & 12.3844       & 1         & 1              & 10.368        & 0.45      \\
		Sim02-19 & 19         & 0.1679                          & 12.6023       & 1         & 1              & 10.5297       & 0.45      \\
		Sim02-20 & 20         & 0.1727                          & 11.9329       & 1         & 1              & 10.0924       & 0.45      \\
		Sim02-21 & 21         & 0.1772                          & 9.301         & 1         & 1              & 10.118        & 0.45      \\
		Sim02-22 & 22         & 0.1815                          & 10.7777       & 1         & 1              & 10.249        & 0.45      \\
		Sim02-23 & 23         & 0.1853                          & 12.9491       & 1         & 1              & 10.134        & 0.45      \\
		Sim02-24 & 24         & 0.1888                          & 10.6571       & 1         & 1              & 10.1151       & 0.45      \\
		Sim02-25 & 25         & 0.1919                          & 9.6297        & 1         & 1              & 9.6608        & 0.45      \\
		Sim02-26 & 26         & 0.1945                          & 8.4331        & 1         & 1              & 8.7821        & 0.45      \\
		Sim02-27 & 27         & 0.1967                          & 10.4921       & 1         & 1              & 9.8995        & 0.45      \\
		Sim02-28 & 28         & 0.1983                          & 11.135        & 1         & 1              & 9.2481        & 0.45      \\
		Sim02-29 & 29         & 0.1994                          & 10.109        & 1         & 1              & 10.4398       & 0.45      \\
		Sim02-30 & 30         & 0.1999                          & 9.5404        & 1         & 1              & 10.1114       & 0.45     
	\end{tabular}
\end{table}

\begin{table}[h]
	\begin{tabular}{c|cllllll}
		Label    & Depth & Density & 210Pb & sd(210Pb) & Thickness & 226Ra   & sd(226Ra) \\
						& (cm) &($g/cm^3$) &(Bq/kg)& & (cm) & (Bq/kg)&\\
		\hline 
		Sim03-01 & 1     & 0.1001  & 6384.1354     & 287.2861  & 1         & 15.8007 & 0.675     \\
		Sim03-02 & 2     & 0.1006  & 3550.0809     & 159.7536  & 1         & 14.5245 & 0.675     \\
		Sim03-03 & 3     & 0.1017  & 1954.5702     & 87.9557   & 1         & 15.6527 & 0.675     \\
		Sim03-04 & 4     & 0.1033  & 1183.8917     & 53.2751   & 1         & 14.5175 & 0.675     \\
		Sim03-05 & 5     & 0.1055  & 760.2132      & 34.2096   & 1         & 14.9242 & 0.675     \\
		Sim03-06 & 6     & 0.1081  & 360.2553      & 16.2115   & 1         & 14.801  & 0.675     \\
		Sim03-07 & 7     & 0.1112  & 212.9402      & 9.5823    & 1         & 14.8738 & 0.675     \\
		Sim03-08 & 8     & 0.1147  & 104.2684      & 4.6921    & 1         & 14.9028 & 0.675     \\
		Sim03-09 & 9     & 0.1185  & 44.3849       & 1.9973    & 1         & 15.0768 & 0.675     \\
		Sim03-10 & 10    & 0.1228  & 18.6447       & 1         & 1         & 15.3764 & 0.675     \\
		Sim03-11 & 11    & 0.1273  & 23.2778       & 1.0475    & 1         & 14.6231 & 0.675     \\
		Sim03-12 & 12    & 0.1321  & 53.1587       & 2.3921    & 1         & 15.1629 & 0.675     \\
		Sim03-13 & 13    & 0.1371  & 97.363        & 4.3813    & 1         & 14.3047 & 0.675     \\
		Sim03-14 & 14    & 0.1422  & 116.9788      & 5.264     & 1         & 14.0261 & 0.675     \\
		Sim03-15 & 15    & 0.1474  & 153.2901      & 6.8981    & 1         & 15.9723 & 0.675     \\
		Sim03-16 & 16    & 0.1526  & 151.8496      & 6.8332    & 1         & 14.7579 & 0.675     \\
		Sim03-17 & 17    & 0.1578  & 136.3609      & 6.1362    & 1         & 16.114  & 0.675     \\
		Sim03-18 & 18    & 0.1629  & 107.2736      & 4.8273    & 1         & 15.4595 & 0.675     \\
		Sim03-19 & 19    & 0.1679  & 76.8966       & 3.4603    & 1         & 15.9439 & 0.675     \\
		Sim03-20 & 20    & 0.1727  & 48.9213       & 2.2015    & 1         & 14.6235 & 0.675     \\
		Sim03-21 & 21    & 0.1772  & 40.4439       & 1.82      & 1         & 14.6716 & 0.675     \\
		Sim03-22 & 22    & 0.1815  & 26.5638       & 1.1954    & 1         & 16.2541 & 0.675     \\
		Sim03-23 & 23    & 0.1853  & 21.714        & 1         & 1         & 14.4826 & 0.675     \\
		Sim03-24 & 24    & 0.1888  & 17.6428       & 1         & 1         & 15.5109 & 0.675     \\
		Sim03-25 & 25    & 0.1919  & 17.3533       & 1         & 1         & 13.6898 & 0.675     \\
		Sim03-26 & 26    & 0.1945  & 17.4211       & 1         & 1         & 14.4684 & 0.675     \\
		Sim03-27 & 27    & 0.1967  & 16.4246       & 1         & 1         & 15.3889 & 0.675     \\
		Sim03-28 & 28    & 0.1983  & 12.4828       & 1         & 1         & 15.0698 & 0.675     \\
		Sim03-29 & 29    & 0.1994  & 13.5514       & 1         & 1         & 15.2346 & 0.675     \\
		Sim03-30 & 30    & 0.1999  & 14.3145       & 1         & 1         & 14.7846 & 0.675    
	\end{tabular}
\end{table}

\end{document}